\documentstyle[12pt]{article}
\begin{document}
\pagestyle{plain}
\setcounter{page}{1}
\baselineskip16pt

\def\equno#1{Eq.~(\ref{#1})}
\def\sectno#1{section~\ref{#1}}
\def\chapno#1{chapter~\ref{#1}}
\def\figno#1{Fig.~(\ref{#1})}

\def\Tr{{\rm Tr}\,}
\def\D#1#2{{\partial #1 \over \partial #2}}
\def\eps{\epsilon}
\def\+{^\dagger}
\def\e{{\rm e}}
\def\gammastr{\gamma_{\rm str}}

\def\ellipsis{~$\ldots$ }
\def\comment#1{}

\begin{titlepage}

\begin{flushright}
PUPT-1479\\
hep-th/9407014
\end{flushright}
\vspace{20 mm}

\begin{center}
{\huge A modified $c=1$ matrix model }

\vspace{5mm}

{\huge with new critical behavior}
\end{center}

\vspace{10 mm}

\begin{center}
{\large Steven S. Gubser and Igor R. Klebanov}

\vspace{3mm}

Joseph Henry Laboratories\\
Princeton University\\
Princeton, New Jersey 08544

\end{center}

\vspace{2cm}

\begin{center}
{\large Abstract}
\end{center}

By introducing a $\int dt \, g\left(\Tr \Phi^2(t)\right)^2$ term into the
action of the $c=1$ matrix model of two-dimensional quantum gravity, we
find a new critical behavior for random surfaces.  The planar limit of the
path integral generates multiple spherical ``bubbles'' which touch one
another at single points.  At a special value of $g$, the sum over
connected surfaces behaves as $\Delta^2 \log\Delta$, where $\Delta$ is the
cosmological constant (the sum over surfaces of area $A$ goes as
$A^{-3}$).  For comparison, in the conventional $c=1$ model the sum over
planar surfaces behaves as $\Delta^2/ \log\Delta$.

\vspace{2cm}
\begin{flushleft}
July 1994
\end{flushleft}
\end{titlepage}
\newpage

\section{Introduction}
\label{Intro}

In recent years a considerable effort has been devoted to understanding the
statistical mechanics of random surfaces.  This problem is relevant to
Polyakov string theory in non-critical dimensions.  Thanks to the matrix
model techniques, we now have a thorough understanding of random surfaces
embedded in one dimension \cite{ceq1,GK,Trieste} or less \cite{clt1}.
Little is known, however, about the physically more interesting higher
dimensional embeddings.  With this in mind, it is important to continue
formulating and solving new matrix models.

One interesting modification of the conventional one-matrix model was solved
in ref.~\cite{Das}.  A new type of critical behavior arises when a term of
the form $g (\Tr \Phi^2)^2$ is added to the action, leading to an integral
over the $N\times N$ hermitian matrix $\Phi$ of the form

\begin{equation}
\log\int {\cal D} \Phi \, \e^{-N \,
   \left[ \Tr \left( {1 \over 2} \Phi^2 -
     {1 \over 4} \lambda \Phi^4 \right) -
    {g \over N} \left( \Tr \Phi^2 \right)^2 \right]} \ .       \label{czero}
\end{equation}

\noindent
The large $N$ limit of this quantity has an interesting geometrical
interpretation.  Feynman graphs of the perturbation theory in $\lambda$
generate the usual connected closed random surfaces, while the
$g (\Tr \Phi^2)^2$ term can glue a pair of such surfaces together at a
point.  This point can be resolved into a tiny neck (a wormhole), so that
the network of such touching surfaces can be assigned an overall genus.  In
the leading large $N$ limit one picks out the surfaces of genus zero, which
look like trees of spherical bubbles such that any two bubbles touch at
most once, and a bubble is not allowed to touch itself.

The authors of ref.~\cite{Das} found a critical line in the $(\lambda, g)$
plane where the free energy \equno{czero} becomes singular.  There exists a
critical value $g_t$ such that, for $g<g_t$, the singularity in the
function of $\lambda$ is characterized by $\gammastr=-1/2$.  In this phase
the touching of random surfaces is irrelevant and one finds the
conventional $c=0$ behavior.  For $g>g_t$, on the other hand,
$\gammastr=1/2$, and one finds branched polymer behavior, which is
dominated by the touching.  Most interestingly, for $g=g_t$, the authors of
ref.~\cite{Das} found a new type of critical behavior with $\gammastr=1/3$.
The interpretation of this is not completely clear, but it has been
suggested that at this point one has an effective theory for random
surfaces embedded in more than one dimension \cite{abc}.  In fact, after
some fine tuning, theories with $\gammastr=1/n$ can be formulated
\cite{Das,abc,korchemsky}.

In view of these results, a continued study of the $g (\Tr \Phi^2)^2$ term
is warranted.  In this paper we investigate its effect on random surfaces
embedded in one dimension.  The relevant model is the matrix quantum
mechanics defined by the path integral

\begin{equation}
{\cal Z} = \int {\cal D} \Phi(t) \, \e^{-N \int_{-T/2}^{T/2} dt \,
   \left[ \Tr \left( {1 \over 2} \dot{\Phi}^2 + {1 \over 2} \Phi^2 -
     {1 \over 4} \lambda \Phi^4 \right) -
    {g \over N} \left( \Tr \Phi^2 \right)^2 \right]} \ ,       \label{PartF}
\end{equation}

\noindent
where $\Phi$ is an $N\times N$ hermitian matrix.  $T$ will be taken to
$\infty$ in the end, but we retain it as a finite quantity for the time
being.  In the large $N$ limit we find a critical line where the free
energy is singular.  As in the model \equno{czero}, there exists a critical
value $g_t$ which separates the conventional matrix model behavior (in this
case $c=1$) from the branched polymer phase.  For $g=g_t$ we find a new
critical behavior:  the sum over connected surfaces of genus zero behaves
as $\Delta^2 \log\Delta$, where $\Delta$ is the cosmological constant.
This should be compared with the usual $c=1$ behavior, $\Delta^2/
\log\Delta$.  Our modified critical behavior is much simpler after we carry
out the inverse Laplace transform:  the sum over surfaces of fixed area $A$
behaves as $A^{-3}$.  The corresponding formula for $c=1$ is more
complicated due to logarithmic corrections.

The organization of this paper is as follows.  In \sectno{Fermion}, we
show how the modified $c=1$ model can be reduced to a fermionic system.  In
\sectno{Phase}, we find conditions on the ground state of the fermionic
hamiltonian which allow us to find the critical points of the model and
evaluate $\gammastr$.  We end with a brief discussion in \sectno{Discuss}.

\section{Reduction to fermions}
\label{Fermion}

The sum over connected surfaces is given, up to a factor, by the free
energy

\begin{equation}
{\cal F} = {1 \over T} \log {\cal Z} \ .                  \label{FreeE}
\end{equation}

\noindent
In the $N \to \infty$ limit, the leading term of $\cal F$ (which is the
only term we will consider in this paper) is $O(N^2)$.  This dominant term
generates touching spherical bubbles embedded in one dimension.  To be more
explicit, the single trace terms in \equno{PartF} generate planar, quartic
Feynman graphs embedded in one dimension:  these are the individual
bubbles, which look just like ordinary $c=1$ surfaces.  The double trace
term, $g \left( \Tr \Phi^2 \right)^2$, gives rise to a pair of ``touching''
propagators.  When these propagators are incorporated into two different
bubbles, we interpret the bubbles as touching.  Arrangements where a pair
of bubbles touch in more than one place, or where a bubble reaches around
and touches itself, are suppressed by a factor $1/N^2$ and thus correspond
to $O(1)$ corrections to $\cal F$.

The path integral in \equno{PartF} can be written as a transition amplitude

\begin{displaymath}
{\cal Z} = \langle f| \e^{-NHT} |i\rangle \qquad {\rm where}
\end{displaymath}

\begin{equation}
H = -{1 \over 2N^2} {\partial^2 \over \partial \Phi^2} +
    \Tr \left( {1 \over 2} \Phi^2 - {1 \over 4} \lambda \Phi^4 \right) -
    {g \over N} \left( \Tr \Phi^2 \right)^2 \ .              \label{MatrixH}
\end{equation}

\noindent
In the $T \to \infty$ limit, $\cal F$ is $-N$ times the ground state energy
of this hamiltonian.  The ground state must be a $SU(N)$-invariant function
of $\Phi$, which is to say a symmetric function of the eigenvalues $x_i$ of
$\Phi$:  dependence on ``angular'' degrees of freedom can only raise the
energy.  As in the $c=1$ model, the key step is to pass to a fermionic
system, in which the ground state is an anti-symmetric function of the
$x_i$.  If $\Delta(x_i)$ is the Vandermonde determinant of the eigenvalues
$x_i$ of $\Phi$, then

\begin{equation}
H = {1 \over \Delta(x_i)} H_f \Delta(x_i)
\end{equation}

\noindent
where $H_f$ is the fermionic hamiltonian:

\begin{equation}
H_f = \sum_{i=1}^N \left( -{1 \over 2N^2} {\partial^2 \over \partial x_i^2}
       + {1 \over 2} x_i^2 - {1 \over 4} \lambda x_i^4 \right)
       - {g \over N} \left( \Tr \Phi^2 \right)^2 \ .         \label{FermionH}
\end{equation}

The $\left( \Tr \Phi^2 \right)^2$ term introduces interactions among the
fermions.  These interactions can be taken care of by a self-consistent
field approach, analogous to Hartree-Fock calculations in multi-electron
atoms.  For the purpose of finding the ground state energy of $H_f$ to
leading order in $N$, it is permissible to make the replacement

\begin{equation}
\left( \Tr \Phi^2 \right)^2 \to
   2 \langle \Tr \Phi^2 \rangle \Tr \Phi^2 -
   \langle \Tr \Phi^2 \rangle^2 \ .                          \label{Replace}
\end{equation}

\noindent
An intuitive way to justify this replacement is to consider $\Tr \Phi^2$ as
varying slightly around its expectation value:  $\Tr \Phi^2 = \langle \Tr
\Phi^2 \rangle + \delta \Tr \Phi^2$.  Expanding $\left( \Tr \Phi^2
\right)^2$ to first order in $\delta \Tr \Phi^2$ yields \equno{Replace}.

Applying \equno{Replace} to \equno{FermionH} turns $H_f$ into a constant
plus a sum of $N$ single-particle hamiltonians:

\begin{equation}
H_f \to gN \left\langle {1 \over N} \Tr \Phi^2 \right\rangle^2 +
\sum_{i=1}^N \left( -{1 \over 2N^2} {\partial^2 \over \partial x_i^2} +
       {1 \over 2} x_i^2 - {1 \over 4} \lambda x_i^4 -
       2g \left\langle {1 \over N} \Tr \Phi^2 \right\rangle x_i^2
       \right) \ .
\end{equation}

\noindent
We write the single-particle hamiltonian as

\begin{equation}
h(x) = -{1 \over 2N^2} {\partial^2 \over \partial x^2} +
       {1 \over 2} (1 - 4gc) x^2 - {1 \over 4} \lambda x^4
    \qquad {\rm where} \qquad
      c = \left\langle {1 \over N} \Tr \Phi^2 \right\rangle \ . \label{Single}
\end{equation}

Since our modified matrix model reduces to free fermions, its solution
proceeds essentially along the lines of the usual $c=1$ solution.  However,
the necessity of imposing a self-consistency condition adds an
extra ingredient which, as we will show, can modify the nature of the
critical behavior.

\section{Leading order solution}
\label{Phase}

To obtain the ground state energy of $H_f$ to leading order in $N$ as $N
\to \infty$, it suffices to use semi-classical techniques.  We regard the
$N$ fermions with single-particle hamiltonian $h(x)$ as a Fermi gas in the
potential $U(x) = {1 \over 2} (1-4gc) x^2 - {1 \over 4} \lambda x^4$.  If
the Fermi energy is $e_F$, then the particle density in $x$ is

\begin{equation}
\rho(x) = {N \over \pi} \sqrt{2 (e_F - U(x))} \ .              \label{RhoX}
\end{equation}

\noindent
Given $\lambda$ and $g$, the values of $e_F$ and $c$ are determined by a
normalization condition and a self-consistency condition:

\begin{equation}
{1 \over N} \int dx \, \rho(x) = 1 \qquad {\rm and} \qquad
  {1 \over N} \int dx \, x^2 \rho(x) = c \ .                   \label{EFandC}
\end{equation}

\noindent
If we let $E$ be $1/N$ times the ground state energy of $H_f$, then

\begin{equation}
-{{\cal F} \over N^2} = E = gc^2 +
e_F - {1 \over 3 \pi} \int dx \,
     \left[ 2(e_F - U(x)) \right]^{3/2} \ .                     \label{GSE}
\end{equation}

\noindent
As a check on the self-consistent field method, one may easily show that

\begin{eqnarray}
\D{E}{\lambda} &=& -{1 \over 4N} \int dx \, x^4 \rho(x)
   = -{1 \over 4} \left\langle {1 \over N} \Tr \Phi^4
        \right\rangle  \nonumber \\
\D{E}{g} &=& -c^2 = -\left\langle {1 \over N} \Tr \Phi^2 \right\rangle^2
   = -\left\langle \left( {1 \over N} \Tr \Phi^2 \right)^2 \right\rangle +
     O(1/N^2) \ .                                               \label{dEdl}
\end{eqnarray}

\noindent
These results agree with what one finds by explicit differentiation of
\equno{PartF} with respect to $\lambda$ and $g$.

At this point it is convenient to rescale $x$ and $h(x)$ as follows:

\begin{displaymath}
y = \sqrt{\lambda \over 1-4gc} x
\end{displaymath}

\begin{equation}
h(y) = {\lambda \over (1-4gc)^2}  h(x)
     = -{1 \over 2 \beta^2} {\partial^2 \over \partial y^2} + V(y)
  \qquad {\rm where}                                          \label{Rescale}
\end{equation}

\begin{displaymath}
\beta = {(1-4gc)^{3/2} \over \lambda} N \qquad {\rm and} \qquad
  V(y) = {1 \over 2} y^2 - {1 \over 4} y^4 \ .
\end{displaymath}

\noindent
We also define a rescaled Fermi energy $\mu_F = {\lambda \over (1-4gc)^2}
e_F$.  The highest possible value for $\mu_F$ is $\mu_c = 1/4$: at this
energy, the fermions completely fill the local well of $V(y)$.

Upon rescaling, \equno{EFandC} becomes

\begin{eqnarray}
{\lambda \over (1-4gc)^{3/2}} &=& {1 \over \pi} \int_{-y_c}^{y_c} dy \,
   \sqrt{2 (\mu_F - V(y))} \equiv I(\mu_F)  \nonumber \\
{\lambda^2 c \over (1-4gc)^{5/2}} &=& {1 \over \pi} \int_{-y_c}^{y_c} dy \,
  y^2 \sqrt{2 (\mu_F - V(y))} \equiv J(\mu_F)                  \label{Permit}
\end{eqnarray}

\noindent
where $y_c = \sqrt{1 - \sqrt{1 - 4 \mu_F}}$.  $c$ can be eliminated from
these equations to yield

\begin{equation}
g = {I^2 \over 4J} \left( {\lambda \over I} \right)^{1/3}
   \left[ 1 - \left( \lambda \over I \right)^{2/3} \right] \ . \label{glCurve}
\end{equation}

\noindent
Eqs.~(\ref{Permit}) and~(\ref{glCurve}) represent necessary conditions on
the ground state, but as we shall see, they are not quite sufficient, as
there are sometimes two possible values of $\mu_F$ for a given $\lambda$
and $g$.  It will not be hard to tell which represents the true ground
state.

\equno{glCurve} determines a family of curves in $\lambda$-$g$ space that
depends on the single parameter $\mu_F$.  \figno{Waterfall} shows twenty of
these curves.  The outer envelope of all the curves is the locus of
critical points where the perturbation series that generates the random
surfaces begins to diverge.  To see this, suppose one varies $\lambda$ and
$g$ from $0$ toward criticality.  As long as the point $(\lambda,g)$
is inside the outer envelope, there is a ground state of $H_f$, and $E$ is
finite.  But as soon as $(\lambda,g)$ crosses the envelope, there is no
solution to \equno{Permit}, so $E$ and hence $\cal F$ are undefined.

There are two regions of the critical curve.  For
$g < g_t = {1 \over 12 \pi} \left( {5 \over 3} \right)^{3/2}$, the curve
given by \equno{glCurve} with $\mu_F = \mu_c$ is the critical curve.  For
$g>g_t$, the critical curve is tangent to a curve given by \equno{glCurve}
with $\mu_F < \mu_c$.  Because of the different critical behaviors we shall
find in these two regions, we shall call the region $g < g_t$ the $c=1$
region and the region $g>g_t$ the branched polymer region.  While the
critical behaviors in these two regions have been found in other models,
right at $g=g_t$ we find something new.

In all the calculations we present, we continue to use the quartic
potential, $\lambda \Tr \Phi^4$, that we started out with in \equno{PartF}.
As a check on universality, however, we have performed the same
calculations with a cubic potential, and we found the same regions of the
critical curve and the same three types of critical behavior.

\subsection{Critical behavior in the $c=1$ region}
\label{c=1}

Let us pick a point $(\lambda_c,g_c)$ on the $c=1$ part of the critical
curve.  To determine the critical behavior of $E$ as $\lambda \to \lambda_c$
with $g$ fixed at $g_c$, it suffices to calculate $\partial E / \partial
\lambda$, which from \equno{dEdl} can be shown to be

\begin{equation}
\D{E}{\lambda} = -{1 \over 4} \left( I \over \lambda \right)^{2/3}
  {K \over I^3} \qquad {\rm where} \qquad
  K = K(\mu_F) \equiv {1 \over \pi} \int_{-y_c}^{y_c} dy \,
    y^4 \sqrt{2 (\mu_F - V(y))} \ .  \label{SimpleE}
\end{equation}

\noindent
To analyze the behavior of $I$, $J$, and $K$ for $\mu_F$ near $\mu_c$,
let us set $\mu_F = \mu_c - \mu$ and write

\begin{equation}
\begin{array}{ccc}
I_c = I(\mu_c) = {2\sqrt{2} \over 3 \pi}\ , &
 J_c = J(\mu_c) = {2\sqrt{2} \over 15 \pi}\ , &
 K_c = K(\mu_c) = {2\sqrt{2} \over 35 \pi}\ ,  \nonumber \\[3pt]
\ I(\mu_c - \mu) = I_c + \delta I \ , &
 \ J(\mu_c - \mu) = J_c + \delta J \ , &
 \ K(\mu_c - \mu) = K_c + \delta K \ .
\end{array}
\end{equation}

\noindent
A standard result from $c=1$ calculations is

\begin{equation}
\D{I}{\mu} = -{1 \over \pi} \int_{-y_c}^{y_c}
  {dy \over \sqrt{2 (\mu_F - V(y))}} = {1 \over \pi\sqrt 2} \log \mu + O(1) \ ,
                                                            \label{LeadingI}
\end{equation}

\noindent
which means $\delta I = {1 \over \pi\sqrt 2}
\mu \log \mu + O(\mu)$.  To calculate
$\delta J$ and $\delta K$, let us first define
$\bar{y}_c = \sqrt{1 + \sqrt{1 - 4\mu_F}}$, so that
$\mu_F - V(y) = {1\over 4}(y_c^2
- y^2) (\bar{y}_c^2 - y^2)$.  Now,

\begin{eqnarray}
\D{J}{\mu} - \D{I}{\mu} &=& {\sqrt{2} \over \pi} \int_{-y_c}^{y_c}
  {(1 - y^2) dy \over \sqrt{(y_c^2 - y^2)(\bar{y}_c^2 - y^2)}}  \nonumber \\
  &\to& {\sqrt{2} \over \pi} \int_{-1}^1 dy = 3 I_c
  \qquad {\rm as} \qquad \mu \to 0  \nonumber \\
\D{K}{\mu} - \D{I}{\mu} &=& {\sqrt{2} \over \pi} \int_{-y_c}^{y_c}
  {(1 - y^4) dy \over \sqrt{(y_c^2 - y^2)(\bar{y}_c^2 - y^2)}}  \nonumber \\
  &\to& {\sqrt{2} \over \pi} \int_{-1}^1 dy \, (1 + y^2) = 4 I_c
  \qquad {\rm as} \qquad \mu \to 0 \ .                      \label{LeadingD}
\end{eqnarray}

\noindent
Keeping terms of order $\mu$ and larger, we find

\begin{equation}
\delta J = \delta I + 3 I_c \mu\qquad {\rm and} \qquad
  \delta K = \delta I + 4 I_c \mu \ .                      \label{LeadingJK}
\end{equation}

Now we perform a perturbation of $\lambda$ away from $\lambda_c$ with $g$
fixed at $g_c$.  If one defines $z = (\lambda / I)^{1/3}$ and $z_c =
(\lambda_c / I_c)^{1/3}$, \equno{glCurve} becomes

\begin{equation}
g {4 J \over I^2} = z (1 - z^2) \ .                     \label{glSimp}
\end{equation}

\noindent
Using this along with \equno{LeadingJK} one can derive

\begin{displaymath}
{\delta z \over z_c} = {1 - z_c^2 \over 1 - 3 z_c^2}
  \left( 3 {\delta I \over I_c} + 15 \mu \right)  \\[3pt]
\end{displaymath}

\begin{equation}
{\delta \lambda \over \lambda_c} = 2 {5 - 6 z_c^2 \over 1 - 3 z_c^2}
  {\delta I \over I_c} + 45 {1 - z_c^2 \over 1 - 3 z_c^2} \mu \ .
                                                         \label{Varyl}
\end{equation}

\noindent
In this and all the rest of our calculations in this section, we retain
terms only up to $O(\mu)$.

Before continuing with the calculation of the critical behavior of $E$, let
us note how \equno{Varyl} can be used to determine the point
$(\lambda_t,g_t)$ where the $\mu_F = \mu_c$ curve ceases to coincide with
the critical line.  \equno{Varyl} actually applies to any point
$(\lambda_c,g_c)$ on the $\mu_F = \mu_c$ curve, except when $1 - 3 z_c^2 =
0$.  If we set $\delta \lambda = 0$, then \equno{Varyl} becomes

\begin{equation}
0 = 2 {5 - 6 z_c^2 \over 1 - 3 z_c^2} {\delta I \over I_c} +
   45 {1 - z_c^2 \over 1 - 3 z_c^2} \mu \ ,              \label{Intersect}
\end{equation}

\noindent
which is a condition on $\lambda_c$ that determines where the $\mu_F =
\mu_c - \mu$ curve intersects the $\mu_F = \mu_c$ curve.  Now,
$(\lambda_t,g_t)$ is the point where the $\mu_F = \mu_c$ curve just starts
to meet curves with lower $\mu_F$.  Hence $\lambda_t$ can be determined by
letting $\mu \to 0$ in \equno{Intersect}.  Since the dominant term on the
right hand side of \equno{Intersect} is the one involving $\delta I / I_c$,
we must have $z_c^2 \to 5/6$, whence $\lambda_t = {1 \over 3 \pi} \left( 5
\over 3 \right)^{3/2}$ and $g_t = {1 \over 12 \pi} \left( {5 \over 3}
\right)^{3/2}$.

Now let us see how $E$ behaves near criticality.  Retaining terms up to
$O(\mu)$ as usual,

\begin{equation}
\D{E}{\lambda} = -{1 \over 4 z^2} {K \over I^3}
  =-{1 \over 4 z_c^2} {K_c \over I_c^3}
    \left[ 1 + \left( {26 \over 3} - 6 {1 - z_c^2 \over 1 - 3 z_c^2} \right)
    {\delta I \over I_c} + \left( {140 \over 3} -
    30 {1 - z_c^2 \over 1 - 3 z_c^2} \right) \mu \right] \ . \label{LeadingE}
\end{equation}

\noindent
Eqs.~(\ref{Varyl}) and~(\ref{LeadingE}) lead to the existence of two
different types of critical behavior.  First, let us take $\lambda_c >
\lambda_t$ $(g_c<g_t)$.  In this case Eqs.~(\ref{Varyl})
and~(\ref{LeadingE}) combine to give

\begin{equation}
\D{E}{\lambda} = -{1 \over 4 z_c^2} {K_c \over I_c^3} \left( 1 +
  {2 \over 3} {2 - 15 z_c^2 \over 5 - 6 z_c^2}
   {\delta \lambda \over \lambda_c} + {70 \over 3}
   {1 - 3 z_c^2 \over 5 - 6 z_c^2} \mu \right)               \label{LE2}
\end{equation}

\noindent
As in the case of ordinary $c=1$, we define the cosmological constant
$\Delta$ by $\lambda = \lambda_c - \Delta$.  To leading order in $\Delta$,

\begin{equation}
\mu = -{\pi \over \sqrt 2} {1 \over z_c^3} {1 - 3 z_c^2 \over 5 - 6 z_c^2}
         {\Delta \over \log \Delta} \ ,                      \label{LeadingM}
\end{equation}

\noindent
which is the behavior typical of ordinary $c=1$.  The leading nonanalytic
behavior of $E$ is

\begin{equation}
E = ({\rm analytic\ in\ } \Delta) - {9 \pi^3 \over 32 \sqrt 2} {1 \over z_c^5}
  \left( {1 - 3 z_c^2 \over 5 - 6 z_c^2} \right)^2 {\Delta^2 \over \log
  \Delta} + ({\rm higher\ order\ terms}) \ ,              \label{UniversalE}
\end{equation}

\noindent
which again is typical of $c=1$. Our conclusion is that, for $g<g_t$,
the touching of random surfaces is irrelevant in the sense that it does not
destroy the $c=1$ behavior.

If we fix $g = g_t$ and let $\lambda \to \lambda_t$, then we get a
different critical behavior: \equno{Varyl} becomes
$\delta \lambda / \lambda_t = -5 \mu$, so that the leading order relation
between $\Delta$ and $\mu$ is just

\begin{equation}
\mu = {\Delta \over 5 \lambda_t} \ .                      \label{NewM}
\end{equation}

\noindent
The fact that $\Delta$ and $\mu$ are simply proportional to each other
is the crucial feature of the new critical behavior.
In contrast to \equno{UniversalE}, the universal part of $E$ is now

\begin{equation}
E = ({\rm analytic\ in\ } \Delta) +
{3^6 \over 2^5 5^4} \left( {3 \over 5} \right)^{1/2}
\pi^3 \Delta^2 \log \Delta +
   ({\rm higher\ order\ terms}) \ .                        \label{NewE}
\end{equation}

Similar calculations are possible for perturbations of $g$ away from
criticality with $\lambda$ held fixed.  If one sets $g = g_c - \Gamma$ and
$\lambda = \lambda_c$, then in the $c=1$ region one finds that the
singularity of $E$ is $\sim -\Gamma^2 / \log \Gamma$.  For
$\lambda = \lambda_t$, $E\sim \Gamma^2 \log \Gamma$ instead.

\subsection{The branched polymer region}
\label{Polymer}

It is clear from \figno{Waterfall} that the region above the curve given by
\equno{glCurve} with $\mu_F = \mu_c$, but below the critical curve, is
covered not once but twice by the family of curves in \equno{glCurve}.
Given $(\lambda,g)$ in this region, there are thus two values of $\mu_F$
and $c$ that satisfy the self-consistent field equations \equno{Permit}.
The true ground state of $H$ corresponds to the solution with the lower
energy $E$.  It is an unsurprising but slightly non-trivial fact that the
solution with lower $E$ is the one with lower $\mu_F$.  At any given point
in \figno{Waterfall} where two curves cross each other, the more
down-sloping curve is the one with lower $\mu_F$, and thus represents the
ground state.

Let us fix $g$ and consider how $\lambda$ varies with $\mu_F$.  The
singularity of the perturbation theory in $\lambda$ occurs at the maximum
of $\lambda(\mu_F)$.  For any $g$, $\lambda (0) =0$.  The behavior of
$\lambda$ with increasing $\mu_F$, however, depends on the value of $g$.

For $g < g_t$, $\lambda$ is an increasing function of $\mu_F$ and the
maximum occurs at the end-point $\mu_F=\mu_c$.  The slope diverges
logarithmically as $\mu_F$ approaches $\mu_c$, which leads to the $c=1$
critical behavior.  For $g= g_t$, $\lambda$ is an increasing function of
$\mu_F$ with a slope that is finite everywhere.  The finiteness of the
slope at $\mu_F=\mu_c$ leads to the modified critical behavior.

For $g > g_t$, $\lambda(\mu_F)$ increases to a quadratic maximum at a value
of $\mu_F$ less than $\mu_c$.  This maximum corresponds to $\lambda$
reaching the critical curve:  $\lambda(\mu_m)=\lambda_c$.  If we increase
$\mu_F$ past $\mu_m$, $\lambda(\mu_F)$ decreases; this region represents
only unphysical solutions to the self-consistent field equations.  In other
words, if for a given value of $\lambda$ there are two possible values of
$\mu_F$, then we pick the smaller one because it is analytically connected
to the perturbative region near $\lambda=0$.  A graph of $\lambda(\mu_F)$
with $g = {5 \over 2} g_t$ is shown in \figno{LamVsMu}.

The main point is that
we can perturb $\lambda$ away from its critical value $\lambda_c$ by
lowering $\mu_F$:  if we write $\lambda = \lambda_c - \Delta$ and $\mu_F =
\mu_m - \mu$, then

\begin{equation}
\Delta = a \mu^2 + ({\rm higher\ order\ terms}) \ ,       \label{QuadraticD}
\end{equation}

\noindent
where $a>0$ is some constant.

{}From \equno{SimpleE} and \equno{QuadraticD} it is simple to show that

\begin{equation}
\D{E}{\lambda} = -{1 \over 4 \lambda_c^{2/3}} \left( {K \over I^{7/3}}
  \Bigg|_{\mu_m} - {1 \over \sqrt{a}} {d \over d \mu_F} {K \over I^{7/3}}
  \Bigg|_{\mu_m} \, \sqrt{\Delta} + O(\Delta) \right) \ .   \label{Taylor}
\end{equation}

\noindent
$K/I^{7/3}$ always has positive derivative, so the universal behavior of
$E$ is

\begin{equation}
E = ({\rm analytic\ in\ } \Delta) -\alpha \Delta^{3/2} +
  ({\rm higher\ order\ terms}) \ ,                         \label{LeadingBPE}
\end{equation}

\noindent
where $\alpha$ is a positive number.
Predictably, the same power law occurs in the universal part of $E$ when we
perturb $g = g_c - \Gamma$ with $\lambda$ fixed at $\lambda_c$: the
leading nonanalytic behavior of $E$ is $\sim -\Gamma^{3/2}$.

It is remarkable that as soon as $g_c$ exceeds $g_t$, the Fermi level $\mu_F$
starts falling.  Since the universal behavior of the $c=1$ model is driven
by the Fermi energy approaching criticality, we regard the dropping of the
Fermi energy as the reason for the transition to the branched polymer
phase.

\section{Discussion}
\label{Discuss}

By performing the inverse Laplace transform of \equno{NewE} we find that,
for $g=g_t$, the sum over connected genus zero surfaces of area $A$ is

\begin{equation}
{\cal F} (A) = {3^6 \over 2^4 5^4} \left( {3 \over 5} \right)^{1/2}
     \pi^3 A^{-3} \ .                                          \label{simps}
\end{equation}

\noindent
This is precisely the KPZ scaling \cite{KPZ} with
$\gammastr=0$. For the conventional $c=1$ matrix model we instead
have \cite{GK}

\begin{equation}
{\cal F} (A) \sim {1 \over A^3 (\log A)^2} \ .
\end{equation}

\noindent
Thus, the critical behavior we have found is actually simpler than the
conventional $c=1$ behavior.  Is it possible that we have penetrated the
$c=1$ barrier, as suggested in ref.~\cite{abc}?  Although we do not have a
definite answer, it seems likely to us that the new critical point can be
described by two-dimensional string theory.  The scaling violations for
$c=1$ have been attributed \cite{JP} to the unusual dependence of the
tachyon potential on the Liouville field, $T(\phi)\sim \phi e^{\phi}$.  It
appears that, if the tachyon potential has the ordinary Liouville form,
$T(\phi)\sim e^{\phi}$, then the simpler scaling of \equno{simps} follows.
Clearly, more work is needed to settle the stringy interpretation of our new
critical theory.

We believe that there are more interesting calculations that can be
performed in the modified $c=1$ matrix model.  Such calculations should
shed light on the critical behavior for higher genus surfaces and,
ultimately, on the double-scaling limit.

\subsection*{Note added}

After this paper was completed, we learned of an interesting paper by
F.~Sugino and O.~Tsuchiya \cite{Sugino} in which results similar to ours
were obtained by means of collective field theory.

\section*{Acknowledgements}

We thank David Gross for interesting discussions.
This work was supported in part by DOE grant DE-FG02-91ER40671,
the NSF Presidential Young Investigator Award PHY-9157482,
James S. McDonnell Foundation grant No. 91-48,
and an A. P. Sloan Foundation Research Fellowship.

\section*{Figures}

\begin{enumerate}

\item The family of curves representing all possible solutions
$(\lambda,g)$ to the self-consistent field equations.  Each curve has
constant Fermi energy $\mu_F$.  The critical curve is the outer envelope of
all these curves.  The transition point from the $c=1$ region to the
branched polymer region is $(\lambda_t,g_t) = \left( {1 \over 3 \pi} \left(
{5 \over 3} \right)^{3/2},{1 \over 12 \pi} \left( {5 \over 3} \right)^{3/2}
\right)$, where curves with different
$\mu_F$ start crossing one another.  When two curves do cross, the one with
the lower $\mu_F$ (or, equivalently, more negative slope) represents the
physical ground state.  Thus, for each curve, only the part to the right of
the contact with the envelope is physical.  \label{Waterfall}

\item A graph of $\lambda$ versus $\mu_F$ for $g$
fixed at ${5\over 2} g_t$.  Only the points with $\mu_F < \mu_m$ represent
physical solutions to the self-consistent field equations.
                                                         \label{LamVsMu}

\end{enumerate}

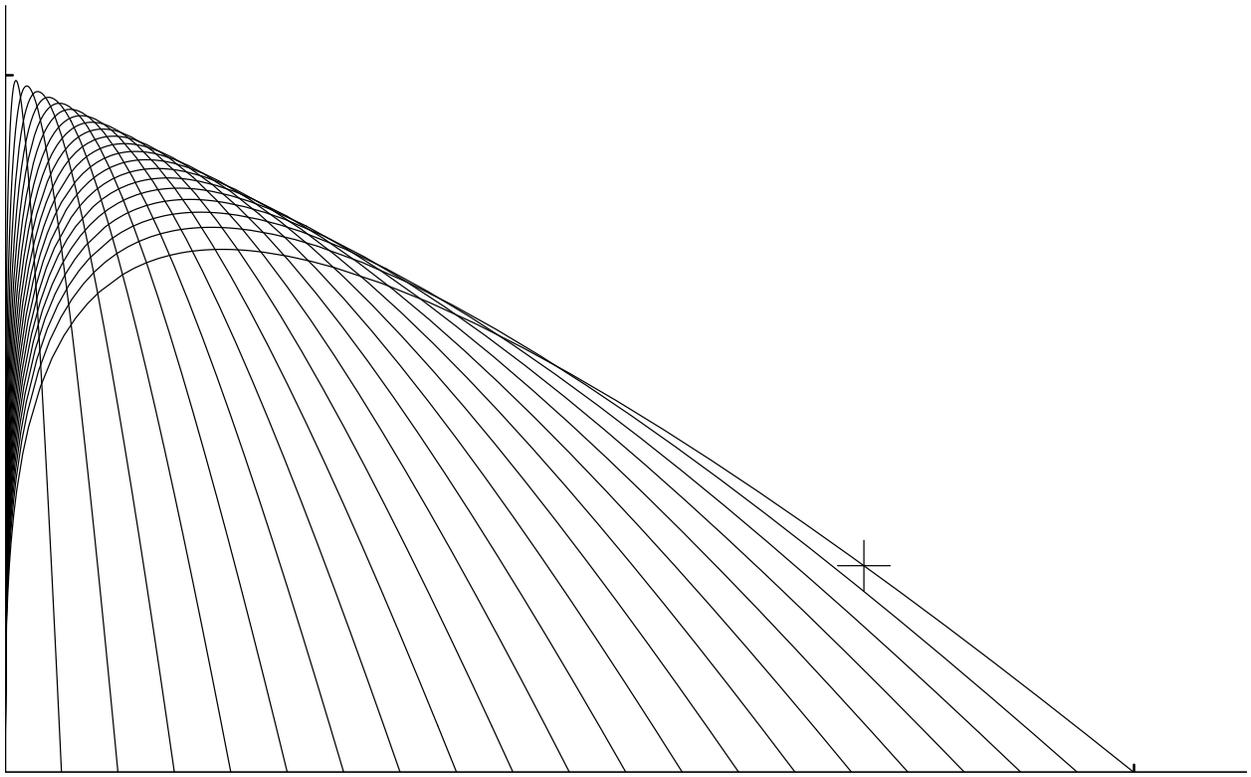
\end{document}